\documentclass[12pt]{iopart}

\usepackage{graphicx}
\renewcommand{\cal}{\mathit}
\newcommand{\cfg}[1]{\ensuremath{\mathcal{C}_{#1}}}
\newcommand{\tm}{\ensuremath{\mathcal{T}}}

\newcommand{\ket}[1]{\ensuremath{| #1 \rangle}}
\newcommand{\braket}[2]{\ensuremath{\langle #1| #2 \rangle}}

\begin{document}

\title{Critical behaviour of the bond-interacting self-avoiding walk}

\author{D P Foster}

\address{Laboratoire de Physique Th\'eorique et Mod\'elisation
(CNRS UMR 8089), Universit\'e de Cergy-Pontoise, 2 ave A. Chauvin
95035 Cergy-Pontoise cedex, France}

\begin{abstract}
The phase diagram for the bond-interacting self-avoiding walk is calculated using transfer matrices on finite strips. The model is  shown to have a richer phase diagram than the related $\Theta$-point model. In addition to the standard collapse transition, we conjecture the existence of a line of transitions in the Berezinskii--Kosterlitz--Thouless (BKT) universality class, terminating in a critical end point. Our results are in contradiction with a previous transfer matrix calculation by Machado K~D, de~Oliveira M~J and Stilck J~F 2001 {\em Phys Rev E\/} {\bf 64}
  051810. 
\end{abstract}


\submitto{\JPA}

\maketitle

\section{Introduction}
Self-avoiding walk models are of interest both as models with which to
understand the thermodynamic behaviour of polymers in solution, and
 as models of theoretical interest in thier own right\cite{Gennes-P-G:1979sh,Cloiseaux:1990mi,Vanderzande:1998ce}. 

The thermodynamic behaviour of a linear polymer in dilute solution is determined by the competition 
between an attractive interaction, due to the difference
between the monomer-monomer affinity and the monomer-solvent affinity,  and an effective repulsive interaction (the excluded-volume interaction) which results from the loss in entropy caused by
 bringing together two segments of polymer.
At high temperature the excluded-volume interaction dominates, leading to the good solvent phase, where the polymer adopts an open configuration, well interpenetrated by the solvent. At low ?temperatures the monomer-monomer affinity dominates and the polymer collapses into a dense ball, and in practice precipitates from solution. This is the bad solvent phase. These two phases are separated by the $\Theta$-point transition\cite{Flory:1971px,Gennes:1972wm}.

The thermodynamics of a polymer in solution is well modelled on the lattice by the so called $\Theta$-point model\cite{Domb:1974cq,Wall:1961tf}, which consists of a self-avoiding walk on a regular lattice with the inclusion of attractive 
interactions between non-consecutively visited nearest-neighbour lattice sites. This model displays behaviour in excellent agreement with experimental results in both two and three dimensions. The critical behaviour
of the ``on-lattice" model is in agreement with the results of equivalent ``off-lattice" models, indicating that the lattice does not affect the critical behaviour. 

The interesting feature of the $\Theta$-point model is that it describes the thermodynamic behaviour of a large class of linear polymers, independently of the detailed chemistry  of the polymer chain. This may be understood as follows: the real polymer chain, if long enough, behaves as a chain of spherical globules, of comparable in size to the persistence length. The detailed chemistry simply defines the size of the globules, and the large scale behaviour of the chain is described by a random walk, constrained by the excluded volume condition (globules repel each other due to  entropic repulsion). The globules interact via effective attractive interactions, incorporating a variety of different microscopic interactions.
 The monomers in the model system then corresponds to a large number of real monomers on the scale of the real polymer chain. Bearing this in mind, when going over to a lattice model, the decision to think of the `monomers' sitting on the lattice sites or 
the lattice bonds is  arbitrary. The choice of placing the interactions between the the bonds would be, in principle, just as reasonable. 

\begin{figure}
\begin{center}
\includegraphics[width=10cm]{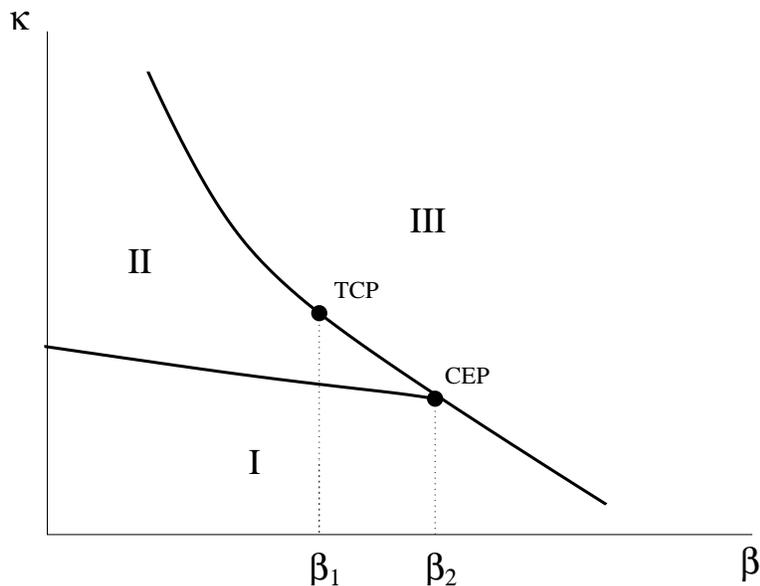}
\end{center}
\caption{Phase diagram proposed by Stilck and co-workers\cite{Stilck:1996rq,Serra:2004qu,
machado:2001aa}. The lower line is a critical line in the self-avoiding walk universality
class, conjectured to terminate 
in a critical end point, {\bf CEP}. The upper line is conjectured to made up of a critical transition line and 
a first order transition line separated by a tricritical point, {\bf TCP}. Phase I is the finite walk phase, phase II an isotropic dense phase, and phase III an anisotropic crystalline phase.}\label{stilckpd}

\end{figure}

\begin{figure}
\begin{center}
\includegraphics[width=10cm,clip]{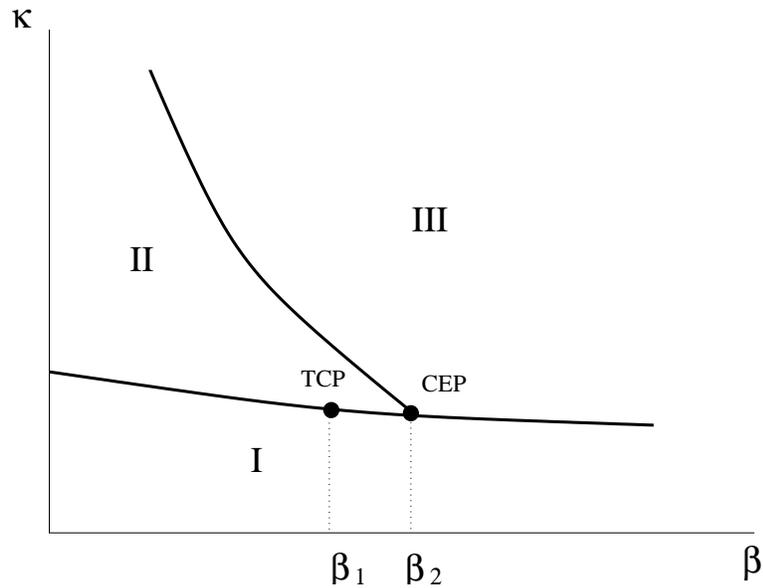}
\end{center}
\caption{A conjectured schematic phase diagram, summarising the results found in this article. The phase diagram splits into three phases: (I) a finite walk phase ($\rho=0$), (II) an isotropic dense walk phase (liquid phase) and  (III) an anisotropic dense walk phase (crystalline phase). 
The transition from phase (I) to phase (II) is critical in the self-avoiding walk universality class for $\beta<\beta_1$ and first-order between $\beta_1$ and
 $\beta_2$ (in analogy with the standard $\Theta$ point model). 
 The transition from phase (I) to phase (III) is first order. 
 The transition occurring at $\beta_1$ is a tricritical point which we conjecture to be
 in the $\Theta$  universality class, whilst the transition at $\beta_2$ is a critical end point. The transition from phase (II) to phase (III) is conjectured to be in the BKT universality class. }\label{pdfinal}
\end{figure}

In this article we present a transfer matrix study of the bond-interacting self-avoiding walk model, and show that the phase diagram differs in many important ways from the standard $\Theta$-point model, 
displaying richer behaviour. This model has been previously studied using extended mean-field type 
calculations \cite{Stilck:1996rq,Serra:2004qu,Buzano:2002hc}. Different methods used by the different authors resulted in two different proposed phase diagrams.
Support for the conjecture due to Stilck and co-workers\cite{Stilck:1996rq,Serra:2004qu} (summarised in
 figure~\ref{stilckpd}) is provided by 
Machado \etal\cite{machado:2001aa}, who studied the model  also using transfer-matrix methods. 
Our results differ in important details, and lend support to the phase diagram proposed by Buzano and Pretti\cite{Buzano:2002hc}. This phase diagram is shown schematically in \fref{pdfinal}. We shall attempt to explain the reasons for these discrepancies.  


\section{The model and the transfer matrix method}

\begin{figure}
\begin{center}
\includegraphics[width=10cm]{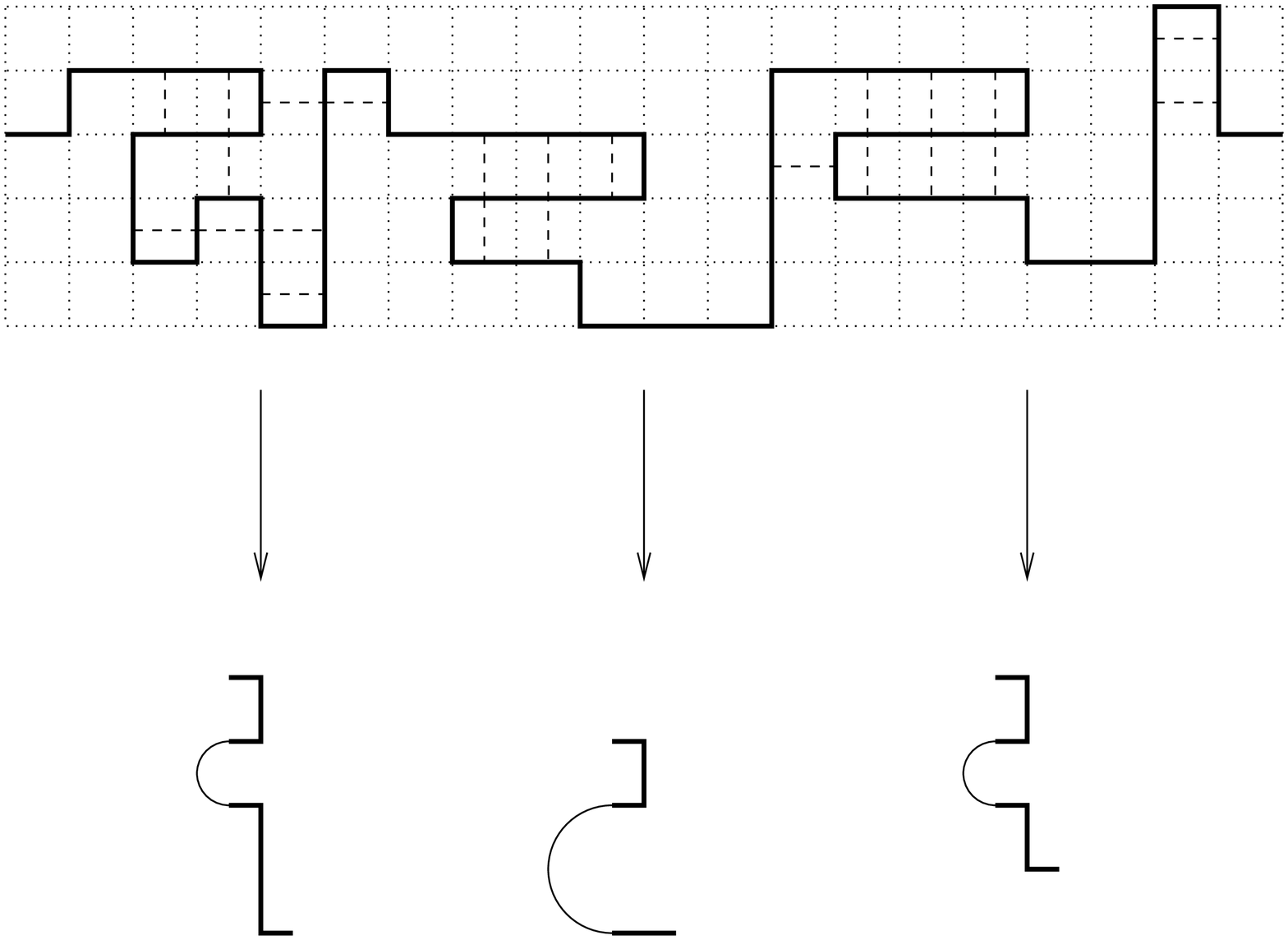}
\end{center}
\caption{Example of a walk configuration, showing example column states. 
The nearest-neighbour interactions are indicated using dashed lines, whilst the thin lines show the connectivities that need to be taken into account in the column state.}\label{ascfg}
\end{figure}

The model studied in this article is the self-avoiding walk on the square lattice. A chemical potential $\mu$, or equivalently a fugacity $\kappa=\exp(-\beta\mu)$, is associated with each step of the walk. An additional attractive energy $-\varepsilon$ is introduced every time two steps of the walk are parallel to each other across the face of a lattice plaquette, see figure~\ref{ascfg}. The thermodynamics of this model may then be studied in the grand-canonical ensemble by introducing the partition function:
\begin{equation}
\mathcal{Z}=\sum_{\rm walks} \kappa^N\exp\left(N_I \beta\varepsilon\right),
\end{equation}
where $N$ is the number of steps in the walk, $N_I$ is the number of interactions, and $\beta=1/kT$ as usual. It is convenient in what follows to set $\varepsilon=1$, which is simply equivalent to a change in
the temperature scale.

Once the partition function has been found, many thermodynamic quantities follow. The (dimensionless) free energy per site is defined as:
\begin{equation}
f=\frac{\log\mathcal{Z}}{\Omega},
\end{equation}
where $\Omega$ denotes the number of sites on the lattice.
The density, $\rho$, of the walk on the lattice and the energy per lattice site follow:
\begin{eqnarray}
\rho&=&\frac{\langle N \rangle}{\Omega}=\kappa\frac{\partial f}{\partial \kappa},\\
e&=&\frac{E}{\Omega}=\frac{\langle N_I\rangle}{\Omega}=\frac{\partial f}{\partial \beta}.
\end{eqnarray}
Further derivatives give the response functions, the susceptibility and the specific heat.

In general there is no exact analytical result for the partition function, and so the thermodynamic quantities must be calculated either by simulation (for example using the Monte-Carlo method) or some approximation scheme must be found enabling the determination of the partition function. In this article we shall use the transfer matrix method, where the exact partition function can be calculated for a lattice of finite width $L$ and of infinite length\cite{Foster:2001xd,Klein:1980jf,Enting:1980kc,Derrida:1981kn,Derrida:1985jw,Derrida:1983tw}. The width of the lattice can then be varied, and the results extrapolated to infinite lattice width.

The standard way of considering the problem is as follows: define the 
restricted 
partition function ${\cal Z}_x(\cfg{0},\cfg{x})$ as the
partition function for a portion of the lattice between $x=0$ and
$x$. The walk has a column state \cfg{0} at the origin and \cfg{x}
in column $x$. One may then write the following recursion
relation:
\begin{equation}\label{recpart}
{\cal Z}_{x+1}(\cfg{0},\cfg{x+1})=\sum_{\cfg{x}} 
{\cal Z}_x(\cfg{0},\cfg{x}){\cal T}(\cfg{x},\cfg{x+1}),
\end{equation}
where $\tm({\cal C}_x,{\cal C}_{x+1})$ is the additional 
Boltzmann weight to add column $x+1$ in configuration ${\cal C}_{x+1}$
next to column $x$ in configuration ${\cal C}_x$. This forms a
(transfer) matrix.

That this recursion should be valid for a spin system is fairly
straightforward, since the interactions are all local. For a polymer
model it is less clear that this should be possible, since one has to
take into account non-local factors, most notably to ensure
that the partition function describes only one chain, without the
formation of ``orphan'' loops. This is done by appropriately defining
the column states ${\cal C}_x$. For the self-avoiding walk problem
with no added interactions (i.e. $\epsilon=0$), it is
sufficient to define a column configuration by the
arrangement of horizontal bonds in the column along with information
about the connectivities between the bonds\cite{Klein:1980jf,Enting:1980kc,Derrida:1981kn}, i.e. information about
which pairs of horizontal bonds are connected by polymer loops to the
left (taking $x$ as increasing towards the right). See \fref{ascfg}
for an example of how to define the column states for the bond-interacting model.  By successive application of \eref{recpart}, one
finds
\begin{equation}
{\cal Z}_x({\cal C}_0,{\cal C}_x)=\langle {\cal C}_0|{\cal T}^x|{\cal
C}_x\rangle.
\end{equation}
Since we are interested in equilibrium thermodynamic behaviour, and we shall be taking the limit $x\to \infty$, the choice of boundary conditions is arbitrary. Choosing periodic boundary conditions in the $x$ direction results in  ${\cal Z}_x=\Tr {\cal T}^x$, giving: 
\begin{equation}
{\cal Z}_x=\sum_i \lambda_i^x
\end{equation} 
terms of the eigenvalues $\lambda_i$ of the transfer matrix $\tm$.
In general the largest eigenvalue  is non-degenerate, and the sum is dominated by this largest eigenvalue, $\lambda_0$, giving, in the limit $x\to\infty$,
\begin{equation}
f=\frac{1}{L}\log\lambda_0,
\end{equation}
where $L$ is the width of the lattice strip.

Having calculated the free energy, it is now possible to calculate the density, the
energy and other local quantities by differentiation. However, since  the eigenvalues of interest will
be determined numerically, it is better to avoid differentiation whenever possible; the eigenvalues themselves may be calculated to arbitrary precision, but the numerical differentiation magnifies round-off errors, leading to a substantial loss of numerical precision. 
This is not a problem, however, since in the transfer matrix framework it  is easy to calculate averages 
over local quantities directly\cite{Foster:2001fi}. 
To see this, it is
necessary first to calculate the probability of having a given configuration
${\cal C}_x$ in column $x$. This probability is simply the ratio of
the partition function restricted to configuration ${\cal C}_x$
and the unrestricted partition function, which in terms of transfer
matrices may be written:
\begin{equation}
p({\cal C}_x)=\lim_{M\to\infty}
\frac{\Tr \left\{T^x | {\cal C}_x \rangle\langle {\cal C}_x |T^{M-x}\right\}}{\Tr T^M},
\end{equation}
where $M$ is the length of the lattice strip.

Writing $|{\cal C}_x \rangle$ in terms of the eigenvectors, $\ket{i}$ of $\tm$ gives:
\begin{equation}
p({\cal C})=\lim_{M\to\infty}
\frac{\sum_i\lambda_i^M\braket{i} {{\cal C}}\braket{{\cal C}}{i}}{\sum_i\lambda_i^M}
\end{equation}

\begin{equation}
p({\cal C})=\braket{0}{{\cal C}}^2,
\end{equation}
where the eigenvectors are normalised. The subscript $x$ may be omitted by invoking translation invariance.

 The density, for example, is then found using
\begin{eqnarray}\nonumber
\rho&=&\sum_{\cal C} \frac{N({\cal C})}{L} p({\cal C}) \\
&=&\sum_{\cal C} \frac{N({\cal C})}{L}\langle 0 | {\cal C} \rangle^2,
\end{eqnarray}
where $N(\mathcal{C})$ is the number of occupied lattice bonds in configuration
 $\mathcal{C}$. The susceptibility can then be calculated either by taking a derivative of the density,
or by calculating directly $\langle N^2\rangle$ {\it for the column}, and hence the fluctuation.  
The two methods give slightly different results for a finite width strip, but agree in the thermodynamic limit. In the present article we choose to calculate the fluctuation directly.
 
It is also possible to calculate two-point  correlation functions, by considering the joint probability
$p({\cal C}_x,{\cal C}_{y})$ of
having a configuration ${\cal C}_x$ in column $x$ and ${\cal C}_{y}$
in column $y$.  The calculation
 proceeds
exactly as above:
\begin{eqnarray}
p({\cal C}_x,{\cal C}_{y})&=&\lim_{M\to\infty}
\frac{
\Tr \left\{ \tm^x | {\cal C}_x \rangle\langle {\cal C}_x |
\tm ^{y-x}| {\cal C}_{y} \rangle \langle {\cal C}_{y} |
\tm^{M-y}\right\}
}{
\Tr \left\{\tm^M\right\}}\\
&=&\lim_{M\to\infty}
 \frac{1}{\lambda_0^M}\langle \mathcal{C}_y|\tm^{M+x-y}|\mathcal{C}_x\rangle\langle\mathcal{C}_x|\tm^{y-x}|\mathcal{C}_y\rangle\\
&\approx& a+b\left(\frac{\lambda_0}{\lambda_1}\right)^{-(y-x)},
\end{eqnarray}
where $a$ and $b$ are constants and $\lambda_1$ is the second largest eigenvalue. 
This leads directly to the identification of the correlation length as\cite{Yeomans:1992zn}
\begin{equation}
\xi=\frac{1}{\log\left(\frac{\lambda_0}{|\lambda_1|}\right)}.
\end{equation}

The correlation length diverges when the two largest eigenvalues become degenerate, signalling the onset of a phase transition. For many spin models there is no phase transition below two dimensions, and so the correlation length should not diverge for finite lattice widths. In the transfer matrix formulation this is seen explicitly by considering the Frobenius-Perron theorem, which states that for a finite-dimensioned matrix with strictly positive entries, the largest eigenvalue is non-degenerate\cite{Horn:1990gc}. In our model, however, there are many zero entries, and the Frobenious-Peron theorem does not apply. The potential degeneracy of the eigenvalues may be used to gain an insight into the phase diagram, and we shall return to this fact below.

The correlation length can be used as the basis of a
 phenomenological renormalisation group analysis\cite{Nightingale:1976fd}. The idea is simple: far from the critical point the correlation length is smaller than the width of the lattice, but as the critical point is approached the divergence of $\xi$ is limited by the smallest dimension, $L$. Since we expect to have scale invariance when both $\xi$ and $L\to\infty$, all thermodynamic quantities should (for sufficiently large lattice sizes) be only by a function of $\xi/L$. The critical point may then be identified with solutions of the equation:
 \begin{equation}\label{night}
 \frac{\xi_L}{L}=\frac{\xi_{L^\prime}}{L^\prime}.
 \end{equation}
 Two lattice sizes are required to determine the estimate, reducing the number of available estimates for extrapolation.
 There is no guarantee that there is a solution to this equation at finite lattice widths, and it is sometimes necessary to show that the gap goes to zero in the thermodynamic limit. 
 There is also no guarantee that a solution to this equation corresponds to a transition point.  Assuming the solution is an estimate for a phase transition, then using the asymptotic behaviour $
 \xi\sim |\kappa-\kappa_c|^{-\nu},
$
 we may infer estimates for $\nu$ at the same time:
 \begin{equation}\label{nuest}
 \frac{1}{\nu_{L,L^\prime}}=\frac{\log\left(\frac{{\rm d}\xi_L}{{d}\kappa}/\frac{{\rm d}\xi_{L^\prime}}{{d}\kappa} \right)}{\log\left(L/L^\prime\right)}-1.
 \end{equation}
 
 Other quantities may be used in a phenomenological renormalisation scheme. For self-avoiding walk type models the density is of particular interest. 
 Close to a critical point  the singular part of the free energy is expected to scale as:
 \begin{equation}
f_s(\kappa,L)=L^{-d}\tilde{f}(|\kappa-\kappa_c|L^{1/\nu}),
 \end{equation}
 where $d$ is the spatial dimension, here $d=2$. Taking the first derivative with respect to $\kappa$ gives:
 \begin{equation}\label{rhoscale}
 \rho_s=L^{1/\nu-2}\tilde{\rho}(|\kappa-\kappa_c|L^{1/\nu}).
 \end{equation}
 We have denoted the density as $\rho_s$ since this is the density related to the singular part of the free energy, which vanishes in the thermodynamic limit. In general there will be a non-zero contribution from the non-singular part of the free energy, giving the limiting density. In walk models, the low-$\kappa$ phase corresponds to finite length walks, and hence the density in the thermodynamic limit is zero. The scaling law then enables us to define a renormalisation scheme\cite{Foster:2003sh,Foster:2003mb}. We start by defining the scaling
 function:
 \begin{equation}\label{phi}
 \varphi_{L,L^\prime}=\frac{\log\left(\rho_s(\kappa,L)/\rho_s(\kappa,L^\prime)\right)}{\log\left(L/L^{\prime})\right)}.
 \end{equation}
Since we wish to look for a family of estimators for the critical point, and taking into account that 
there are potentially severe parity effects, it is convenient to set $L^\prime=L-2$. Estimates of the critical point are then found by looking for solutions of the equation:
\begin{equation}\label{phisol}
\varphi_{L,L-2}(\kappa^*_L)=\varphi_{L-2,L-4}(\kappa^*_L).
\end{equation}
If such a solution exists, then $\kappa_c=\lim_{L\to\infty} \kappa_L^*$ and $\nu=\lim_{L\to \infty}(1/\left(2+\varphi_{L,L-2}(\kappa_L^*)\right)$.
Other quantities will be used later, but the idea remains the same.

Of course estimates of transition points can also be found by looking at appropriate susceptibilities, and identifying the transition with the peak of the function. 

Each method has its advantages and disadvantages. The phenomenological renormalisation group method proposed by Nightingale is usually the most asymptotic, but we shall argue here that some of the lines found using this method do not correspond to transition lines, most notably that found by Machado \etal\cite{machado:2001aa}. Caution is therefore important, and it is wise to crosscheck results using different methods.

\section{The $\kappa$--$\beta$ Phase Diagram}

In this section, the phase diagram is mapped out in the $\kappa$--$\beta$ plane.

It is expected that the model will have a behaviour similar to the $\Theta$-point model as $\kappa$ is increased from zero: at low enough $\kappa$ the length of the walk is finite, and since the walk sits on an infinite lattice, the density is zero. The average length of the walk diverges along the line 
$\kappa=\kappa_1(\beta)$ , either smoothly, for small enough $\beta$, or discontinuously, for large enough $\beta$. In the first case we have a critical phase transition in the same universality class as the self-avoiding walk model ($\beta=0$) and in the second we have a first order transition. In the $\Theta$ point model the point separating the two regimes is a tricritical point, with a correlation length exponent 
$\nu=4/7$, different from the self-avoiding walk value $\nu=3/4$.
Similarly, as $\beta$ is increased along the line $\kappa=\kappa_1(\beta)$, the interactions of the bond-interacting self-avoiding walk will drive a collapse transition. 
 In the $\Theta$-point model there are no other phase transitions, but for the bond-interacting model
extended mean-field calculations predict a phase transition between two dense walk phases\cite{Stilck:1996rq,Serra:2004qu,Buzano:2002hc}, 
an isotropic phase (the liquid phase II in \fref{pdfinal}) and an anisotropic phase (the crystalline phase
 III in \fref{pdfinal}). We will identify this transition line, $\kappa=\kappa_2(\beta)$, with a crystallisation transition. This line will terminate on the  line $\kappa=\kappa_1(\beta)$ at a value of $\beta=\beta_2$.
 An important question to be answered is whether $\beta_1=\beta_2$, i.e., does the walk, as the line $\kappa=\kappa_1(\beta)$ is followed, collapse then crystallise, or rather collapse directly to a crystalline, anisotropic, state? The results we present bellow tend to support the first scenario. 

As mentioned in the previous section, the correlation length diverges when the two largest eigenvalues become degenerate. As a result of the connectivity constraints, this may occur in the walk models at finite lattice width. In the low-$\kappa$ phase the density is zero, and the corresponding largest eigenvalue is equal to one for all values of $\kappa$. The second largest eigenvalue steadily increases as $\kappa$ increases, and for some value of  $\kappa=\kappa_L^*(\beta)$ the two eigenvalues become degenerate.
 For $\kappa>\kappa_c$ the walk will fill  the lattice with a finite density. The largest eigenvalue is now bigger than one, and clearly does not correspond to the vacuum state. An estimate for the location of 
the transition, $\kappa=\kappa_1(\beta)$, is then determined by equating the two largest eigenvalues $\lambda_0=\lambda_1=1$. 
 
 It turns out that the transfer matrix block-diagonalises naturally into three sectors: the vacuum state forms a block on its own, an odd sector (corresponding to walks with an odd number of links per column) and an even sector (where the walks have an even number of links per column). Since the vacuum state is trivial, and the eigenvalue is constant and equal to one, no calculation is required. Usually the calculation of the transfer matrix is limited to the odd sector and then the transition  point is determined by simply setting $\lambda_1=1$\cite{Derrida:1985jw}. Since the smallest 
 possible increase of density on a finite width lattice is $\Delta\rho=1/L$, this transition, which exists at finite lattice widths, will appear as a first-order transition, becoming critical in the infinite lattice width limit. 

This method is easy to implement, and often a good choice to find estimates of the critical point, since we find one estimate per lattice width $L$. Since the number of lattice widths one can calculate is limited by the rapid increase in the size of the transfer matrix, it is important to use estimators that require the smallest number of lattice sizes if we are to  apply a reasonable extrapolation scheme. The disadvantage of this method is that the estimates of the critical point may start reasonably far from the true critical point.

In what follows, we define $\lambda_1$ as the largest eigenvalue of the
odd sector and $\lambda_2$ as the largest eigenvalue of the even sector.

\begin{figure}
\begin{center}
\includegraphics[width=10cm,clip]{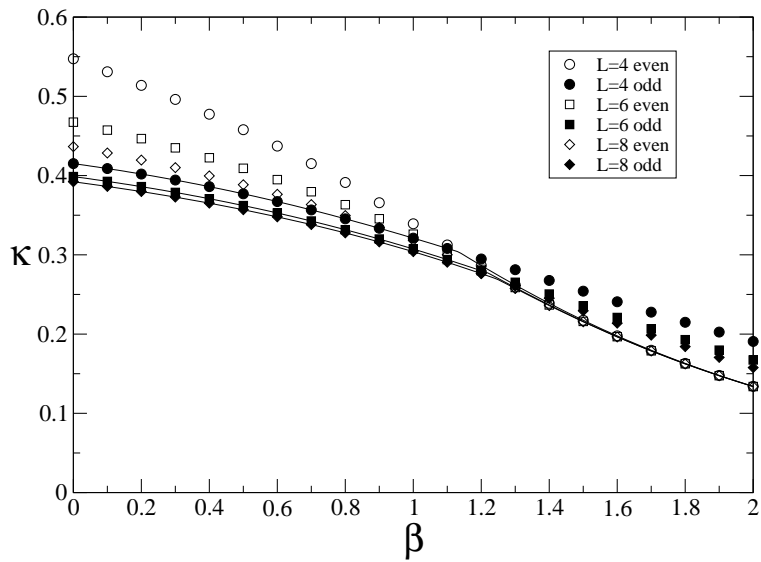}
\end{center}
\caption{Estimates of the zero-density/finite-density phase transition line determined by setting $\lambda=1$ for the even sector (open symbols) and odd sector (closed symbols). The solid lines indicate the transition line estimates. The collapse transition is identified with the point $\lambda_1=\lambda_2=1$.}\label{evone}
\end{figure}

The even sector of the matrix corresponds to walk configurations that cross the lattice an even number of times. The whole analysis may be repeated with the even sector, and it would be expected that the same final results would be found, the difference between the walk crossing once or twice simply corresponding to a change in boundary conditions. For the standard $\Theta$-point model, $\lambda_1>\lambda_2$ everywhere, and so the lines found are simply less asymptotic estimates. 
This is no longer true for the bond-interacting self-avoiding walk {\em for even lattice sizes}, 
where for high enough $\beta$, $\lambda_2$ reaches one first as $\kappa$ increases. 
This is shown in \fref{evone}. There is a
point, then, on the $\kappa=\kappa_1(\beta)$ line in which we simultaneously have $\lambda_1=\lambda_2=1$.
We identify this point with the crystallisation transition. This is a reasonable identification if one considers that in the crystalline phase the walk is trying to align its bonds with a lattice direction, and so 
must minimise the number of corners. For {\em even} lattice sizes the easiest way of achieving this is for the walk to fold back onto itself an even number of times. The corresponding configurations sit in the even sector of the transfer matrix, leading to the observed crossover in the eigenvalues. 
This is of course not the case for odd lattice widths, where the same argument 
leads to an odd number of folds, with corresponding configurations which sit in the odd sector
 leading to no such degeneracy in  eigenvalues, i.e. $\lambda_1>\lambda_2$ everywhere.

\begin{figure}
\begin{center}
\includegraphics[width=10cm,clip]{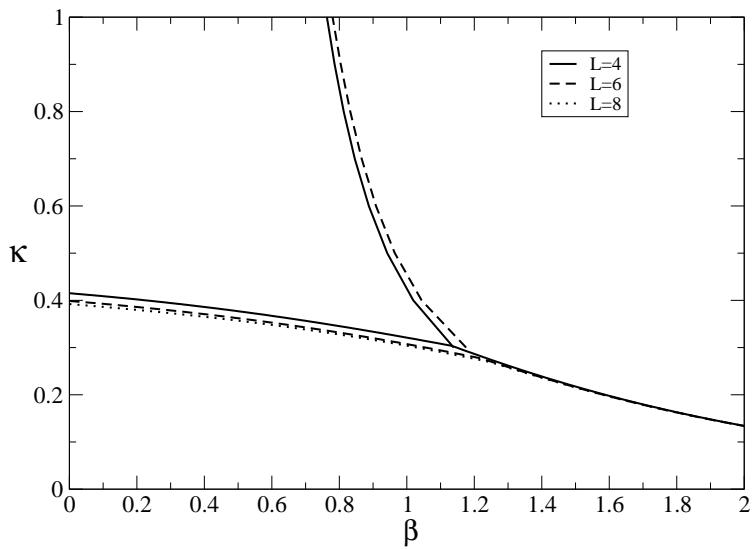}
\end{center}
\caption{Complete phase diagram found using eigenvalue crossings. The lower line is found using the condition $\max(\lambda_1,\lambda_2)=1$ whilst the upper line is found using the condition $\lambda_1=\lambda_2$.}\label{lampd}
\end{figure}

By extension, the condition $\lambda_1=\lambda_2$ defines a line in the $\kappa$---$\beta$ plane. When $\lambda_1=\lambda_2<1$, then the largest eigenvalue corresponds to the vacuum state $\lambda_0=1$. Crossings of subdominant eigenvalues indicate a change of local order, and define  
so-called disorder lines\cite{Stephenson:1970yj,Stephenson:1970ew,Stephenson:1970md,Stephenson:1969qd}. However, when $\lambda_1=\lambda_2>1$ we have a crossing of the largest and second largest eigenvalues, and thus have a divergent correlation length. This condition defines a transition line, possibly critical, in the dense walk phase. We identify this line with the crystallisation transition line $\kappa=\kappa_2(\beta)$. 
The full phase diagram found using these arguments is shown in figure~\ref{lampd}.

\begin{figure}
\begin{center}
\includegraphics[width=10cm,clip]{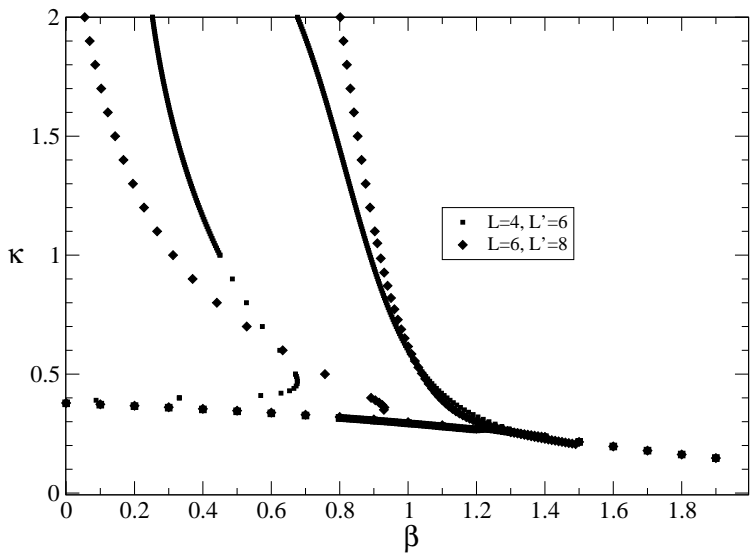}
\end{center}
\caption{Solutions of the renormalisation equation \eref{night}. The left-most line is consistent with the transition line conjectured in Machado \etal\cite{machado:2001aa}, whilst the right-most line is consistent with the transition line found using the condition $\lambda_1=\lambda_2$, and shown in figure~\ref{lampd}.
The question remains as to which lines are true transition lines (see text).}\label{nightpd1}
\end{figure}

Machado \etal\cite{machado:2001aa} performed a transfer matrix calculation
for this model, and found evidence of a transition line, though at a much lower value of $\beta$ than the transition line predicted here.
In order to resolve this discrepancy, we look for solutions to the phenomenological renormalisation group equation \eref{night}. Results for even lattice sizes are shown in \fref{nightpd1}. In determining the correlation lengths, best results are found if the two largest eigenvalues are used. To be sure of having the two largest eigenvalues, it is necessary to calculate in both the odd and the even sectors. It turns out that in the region of interest the two largest eigenvalues appear in different sectors, which simplifies their calculation. 
These results
are complementary to those given by Machado \etal\cite{machado:2001aa}, who limited their study to the odd sector of the transfer matrix, and mostly to odd lattice sizes.

In figure~\ref{nightpd1} we see the appearance of an additional line, which is consistent with the transition line proposed by Machado \etal\cite{machado:2001aa}.  

\begin{figure}
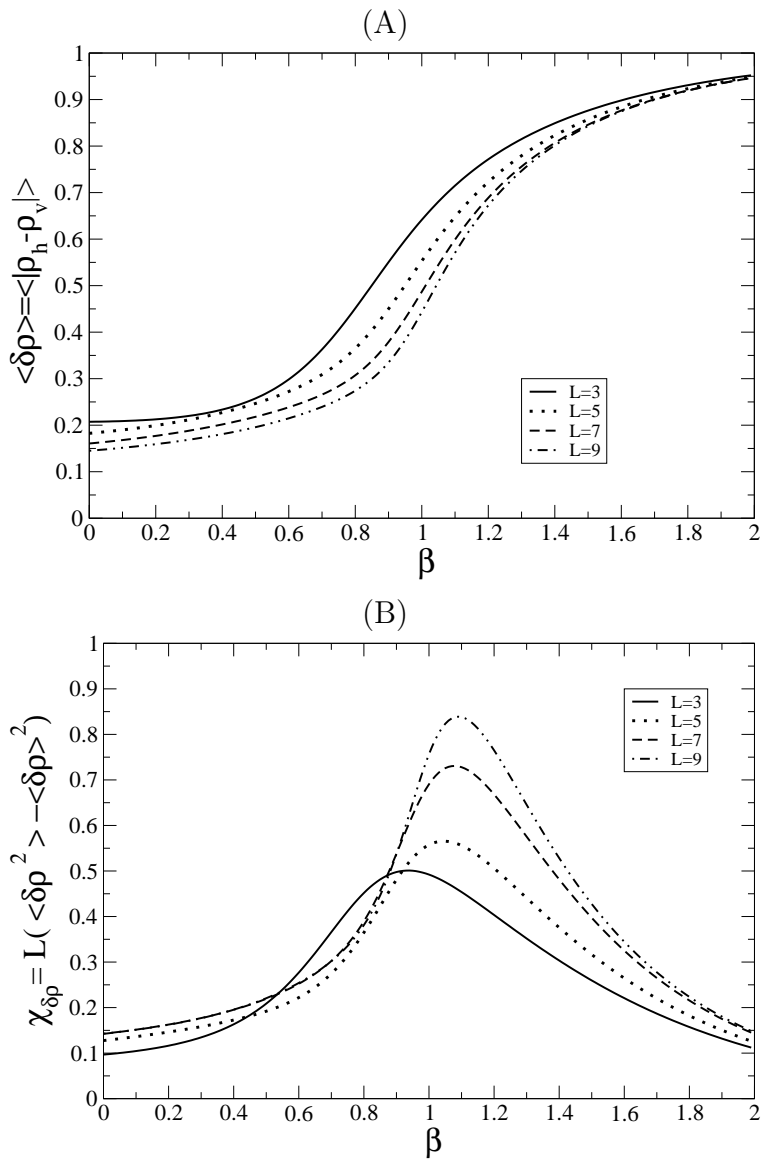

\begin{center}
(A)

\includegraphics[width=10cm,clip]{dr.eps}

(B)

\includegraphics[width=10cm,clip]{fluctdr.eps}
\end{center}
\caption{The order parameter,
$\langle\rho_h-\rho_v|\rangle$,  suitable for the high-$\kappa$ transition 
(A) and its fluctuations (B) for $\kappa=0.5$.}\label{deltarho}
\end{figure}

\begin{figure}
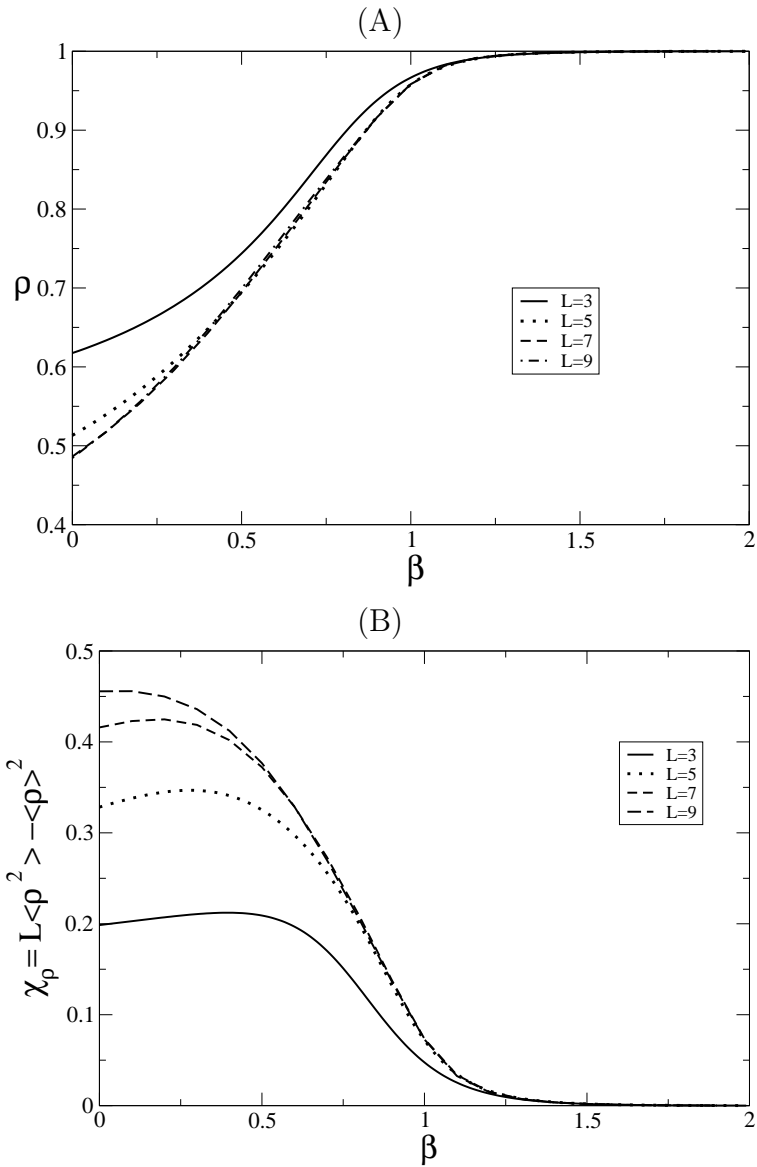

\begin{center}
(A)

\includegraphics[width=10cm,clip]{rho.eps}

(B)

\includegraphics[width=10cm,clip]{fluctrho.eps}
\end{center}
\caption{The density, $\rho$, (A) and its fluctuations (B) for $\kappa=0.5$.
}\label{rho}
\end{figure}

\begin{figure}
\begin{center}
\includegraphics[width=10cm,clip]{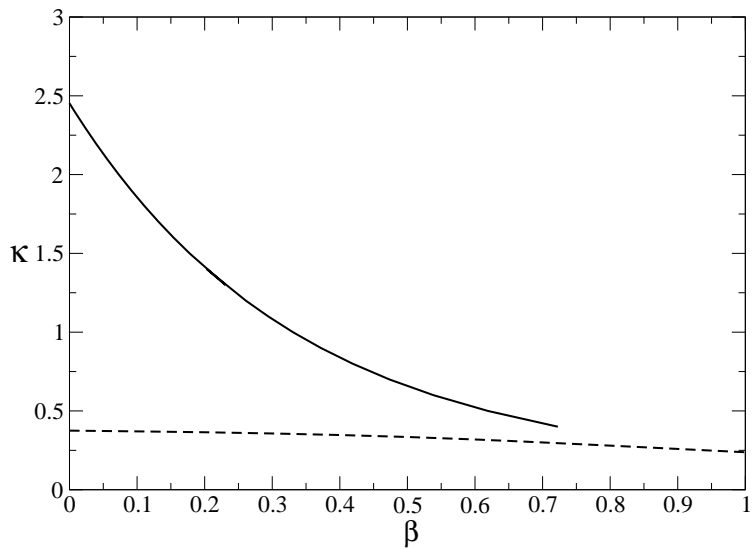}
\end{center}
\caption{Solutions of the renormalisation equation~\eref{night} for lattice sizes $L=3, L^\prime=5$ 
for the $\Theta$-point model, showing the existence of a high-density solution, which does not correspond to a known transition line. The dashed line corresponds to the transition line,  and the solid line to the high density solution. The region where it may be expected to join the true transition line is consistent with the finite-size estimate for the $\Theta$-point. This seems to remain true as the lattice width is increased.}\label{thetaline}
\end{figure}

The question now arises: which lines correspond to true phase transitions? It is natural to look for indicators such as response functions (fluctuations in the order parameter, specific heats, etc.), 
which may be expected to show a singularity at the phase transition.
A first step is to define a suitable order parameter; a good choice is the difference in densities 
of horizontal and vertical bonds, $\delta\rho=|\rho_h-\rho_v|$. The energy is maximised when all the monomers are parallel and pointing in the same direction. The effect of increasing temperature is to disorder this ground-state, mixing up the two phases, and restoring isotropy. Figure~\ref{deltarho} shows both $\delta\rho$, and corresponding fluctuations, for $\kappa=0.5$, i.e. within the dense walk region of the phase diagram. 
The response function is  $\chi_{\delta\rho}=L(\langle \delta\rho^2\rangle-\langle \delta\rho\rangle^2)$. Figure~\ref{rho}, on the other hand, shows the density and its corresponding fluctuations. The density of interactions is directly related to the density, and gives similar results. It is clear from these figures that there is a transition at the higher value of $\beta$. We maintain that the lower-$\beta$ line does not correspond to a thermodynamic phase transition.
 Indeed a similar line exists in the $\Theta$ model, shown in figure~\ref{thetaline}, whilst it is known that there is no high-$\kappa$ transition line.
 It is interesting to note that the line does, however, seem to join up with the low-$\kappa$ line
 at the location of the tricritical $\Theta$-point. This is particularly interesting, since the equivalent line
 in the bond-interacting model also joins the line $\kappa=\kappa_1(\beta)$ 
 at a point consistent with the collapse transition (figure~\ref{nightpd1}), as noted by Machado \etal\cite{machado:2001aa}. 
 It is less clear is whether this point is the same as 
 the crystallisation point found by the condition $\lambda_1=\lambda_2=1$, i.e. the termination point of the line $\kappa=\kappa_2(\beta)$. We shall return to this issue later.  

\begin{figure}
\begin{center}
\includegraphics[width=10cm,clip]{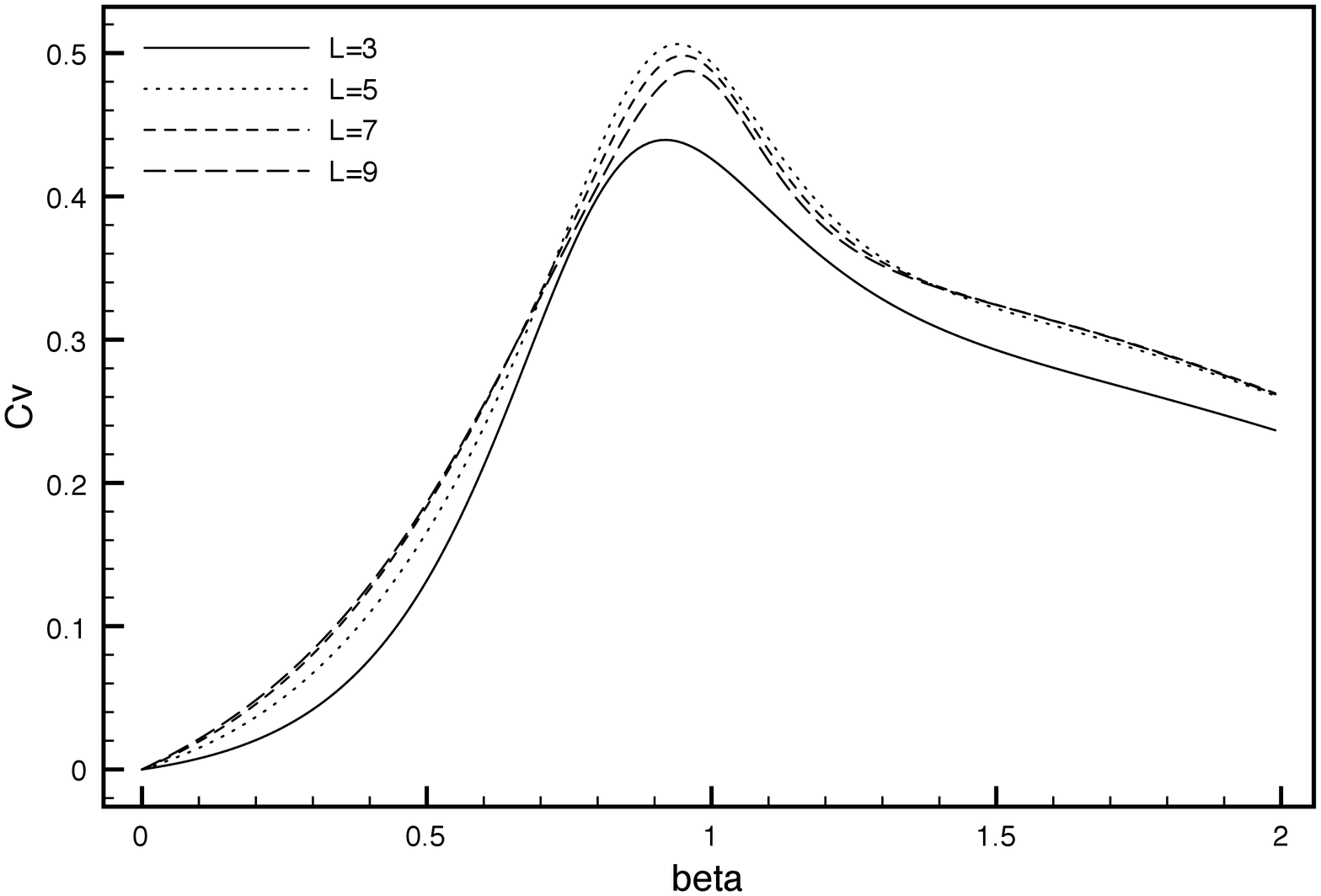}
\end{center}
\caption{Plots for the specific heat for $\kappa=0.5$. The peaks of the specific heat converge to a finite value, at a value of $\beta$ lower than the phase transition. The shape and behaviour are reminiscent of
a KT type transition.}\label{cv}
\end{figure}

In figure~\ref{cv}
the specific heat is plotted for $\kappa=0.5$, 
calculated directly from the grand-canonical free energy using the fluctuation-dissipation theory:
\begin{equation}
C_v={L\beta^2}\left(\langle (e-\mu \rho)^2\rangle-\langle e- \mu \rho\rangle^2\right),
\end{equation}
where $e$ is the energy per site and $\mu$ is the chemical potential ($\beta\mu=-\log\kappa$).

The specific heat does not diverge, and has a broad peak, located at a lower value of $\beta$ than 
the transition line for fixed $\kappa$. This is reminiscent of the Berezinskii-Kosterlitz-Thouless (BKT)
 transition\cite{Nelson:1983be}, which is an infinite ordered phase transition, i.e. none of the derivatives of the free energy are singular. This identification is reinforced by the observation that the order parameter, $\delta \rho$, is not related to a derivative
of the free energy. It is known that walk models with geometrically frustrated interactions show BKT type transitions in the limit $\kappa\to \infty$\cite{Saleur:1986on}. To the best of our knowledge, 
this would be the first time 
this has been observed at finite $\kappa$.
 
\begin{figure}
\begin{center}
\includegraphics[width=10cm,clip]{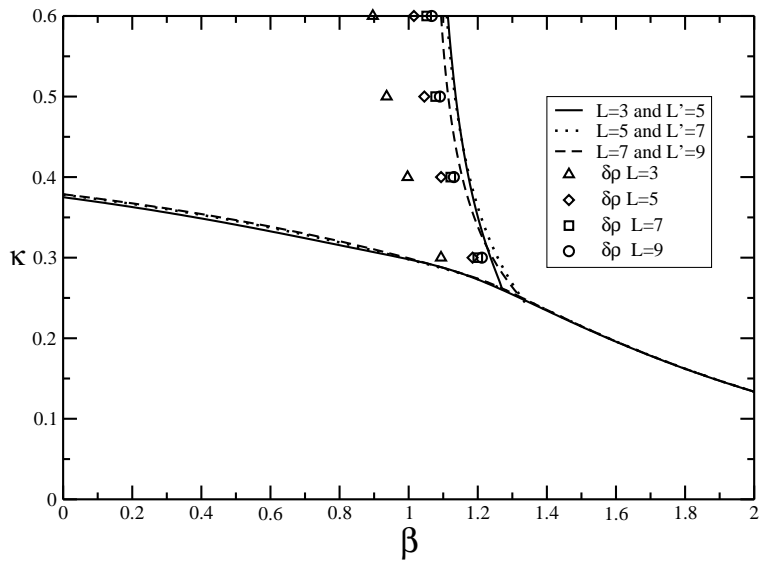}
\end{center}
\caption{Proposed phase diagram calculated using the phenomenological renormalisation group
equation~\eref{night} using  using odd lattice widths. On the same diagram we plot the peaks of $\chi_{\delta\rho}$.}\label{nightpd}
\end{figure}

Figure~\ref{nightpd} shows the phase diagram calculated using Nightingale's phenomenological renormalisation group method for odd lattice sizes. On the same diagram we have plotted the peaks of 
$\chi_{\delta\rho}$, which may be seen to converge nicely to the same line. These results agree with those found earlier for even lattice sizes, using the eigenvalue crossing method (see figure~\ref{lampd}).

\section{Locating the multicritical points}

In this section we shall try to pinpoint the location of the collapse transition and  the crystallisation transition along the $\kappa=\kappa_1(\beta)$ line in order to 
 determine if these points are the same ($\beta_1=\beta_2$), or if there are two distinct transitions
 ($\beta_1\neq\beta_2$).

The most common way of locating the collapse transition 
is to plot $\nu$ estimates. For $T>T_{coll}$, the collapse-transition temperature, the estimates tend to the self-avoiding walk value, whilst for $T<T_{coll}$ the estimates of $\nu$ are expected to tend to 1/2; as $\kappa\to\kappa^*$, the correlation length may be identified with the linear size occupied by the walk. Since, at the first order transition, the walk fills the lattice to a finite density, its linear dimension scales as $N^{1/2}$, or $|\kappa-\kappa^*|^{-1/2}$. The value of $\nu$ at the collapse transition is expected to take on an intermediate value. Crossings would be expected to converge to 
the point $(\beta_{coll},\nu_{coll})$ at the transition point. Additionally, \eref{nuest} is simply a phenomenological renormalisation group equation, and so crossings of $\nu$ may be identified with fixed points, corresponding to different universality classes.

\begin{figure}
\begin{center}
\includegraphics[width=10cm,clip]{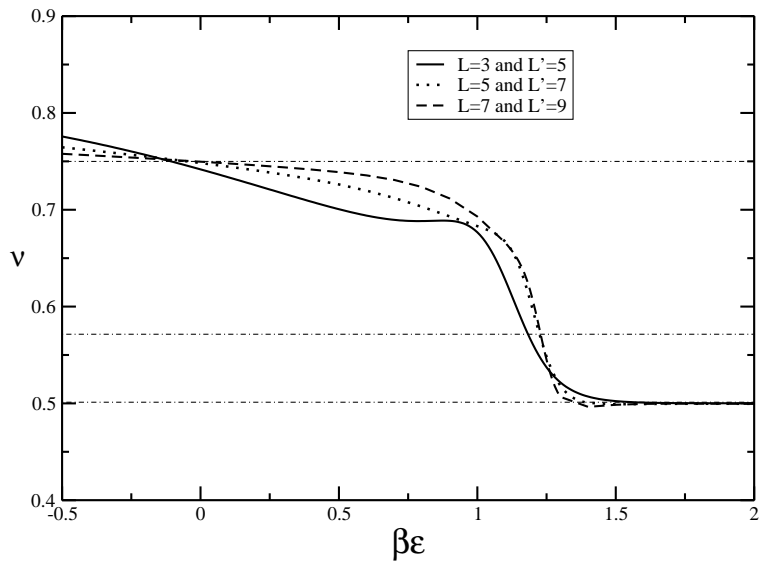}
\end{center}
\caption{Estimates of $\nu$ calculated using \eref{nuest}.  Using phenomenological renormalisation arguments, crossings may be identified with fixed points. Dashed lines indicate the three special values of $\nu$ for the $\Theta$-point model: $\nu_{SAW}=3/4$, $\nu_{\Theta}=4/7$ and $\nu_{\rm dense}=0.5$. We clearly see a fixed point corresponding to the pure self-avoiding walk model behaviour, as well as the correct dense walk limit as $\beta$ is increased. Whilst the crossing between the $L=3, L^\prime=5$ and the $L=5, L^\prime=7$ line is far from $\nu=4/7$, 
the crossing between the $L=5, L^\prime=7$ and $L=7, L^\prime=9$ lines is rather close. 
Unfortunately, with so few estimates, it is not possible to extrapolate. Note the possible appearance of an additional solution with $\nu\approx 2/3$, see text for discussion.}\label{nuplots}
\end{figure}

Figure~\ref{nuplots} shows the estimates for $\nu$ calculated using Nightingale phenomenological renormalisation, see~\eref{nuest}, for the odd lattice sizes, for which $\lambda_1>\lambda_2$ along the whole line. The estimates may clearly be seen to cross. There is also 
an intriguing series of shoulders, which  seem to accumulate to form a fixed point with a value of $\nu\approx 2/3$, close
to the branched polymer value for $\nu$\cite{Parisi:1981kv}. This may be an indication that the walk forms branching structures before fully collapsing. 
This is  more likely to be a crossover effect than a true phase transition, but this point warrants further investigation.

The values calculated for the crossing points are given in tables~\ref{values} and~\ref{valuesb}. With so few data points it is not possible to extrapolate, particularly remembering that there are strong parity effects, however, looking at the odd lattice sizes, the transition is likely to be in the region of $\beta_{\rm coll}\approx 1.22-1.24$. Values of $\nu$ are consistent with a collapse transition in the $\Theta$-point universality class ($\nu_{3,5,7}=0.525974$, $\nu_{5,7,9}=0.556653$ and 
$\nu_{\Theta}=4/7=0.5714286$).

\begin{figure}
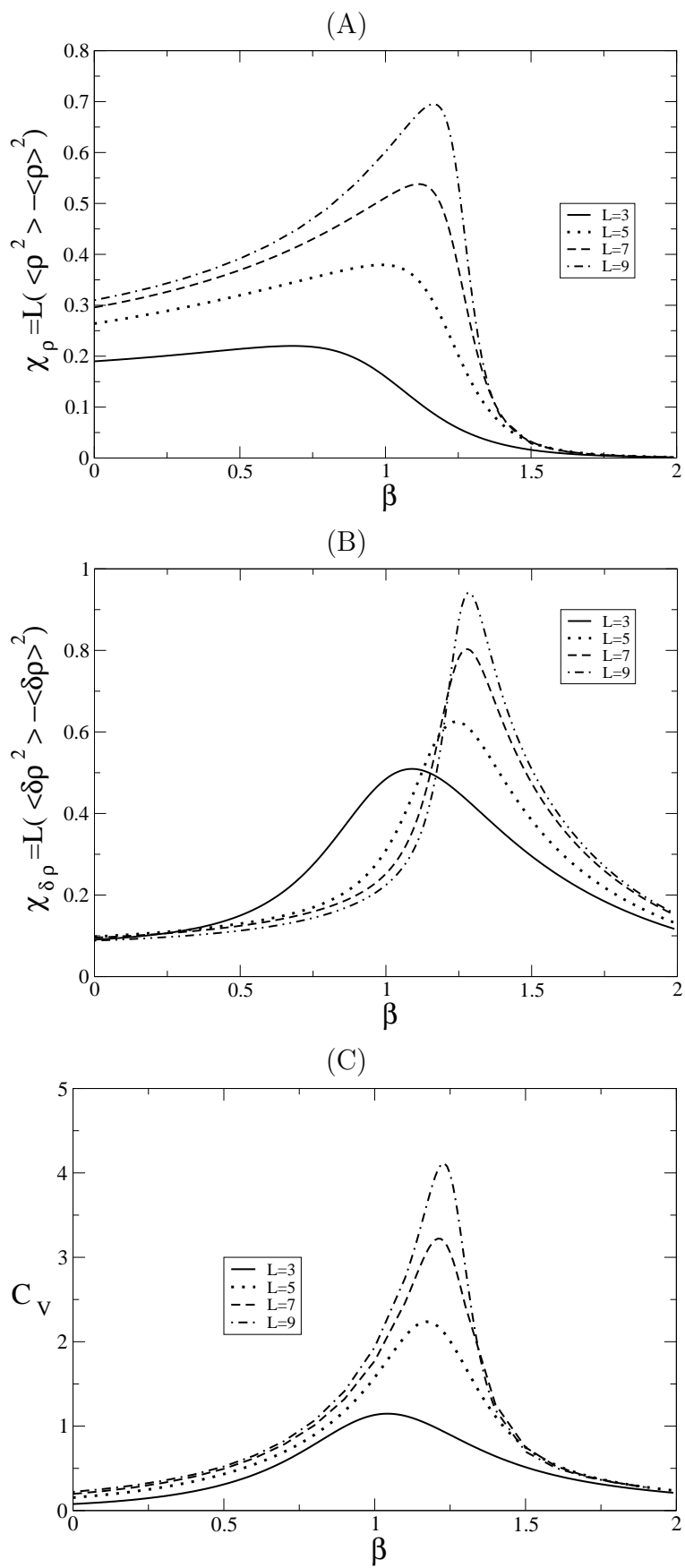

\begin{center}
(A)

\includegraphics[width=10cm,clip]{fluctrkc}

(B)

\includegraphics[width=10cm,clip]{fluctdrkc}

(C)

\includegraphics[width=10cm,clip]{cvkc}
\end{center}
\caption{Plots of (A) $\chi_\rho$, (B) $\chi_{\delta\rho}$ and (C) the specific heat with $\kappa=\kappa^*{\beta}$, along the low-$\kappa$-transition line.}\label{chiplots}
\end{figure}

The location of the collapse transition can be estimated from the peaks of the response function for the corresponding order parameter. The density provides a good order parameter for the collapse transition, whilst $\delta\rho$ provides a good order parameter for the crystallisation transition. 
  The plots of $\chi_\rho$ and $\chi_{\delta\rho}$ along the transition line are given in figure~\ref{chiplots}, and the location of the corresponding peaks is given in tables~\ref{values} and~\ref{valuesb}.  The results derived from $\chi_{\delta\rho}$  indicate a transition which is certainly at a value of $\beta>1.27$ whilst $\chi_\rho$ indicates that the collapse transition occurs at a  
  value of $\beta<1.25$. This leads to the  conclusion that there are at least two distinct multicritical 
  points along the low-$\kappa$ transition line ($\kappa=\kappa_1(\beta)$. 
 
\begin{figure}
\begin{center}
\includegraphics[width=10cm,clip]{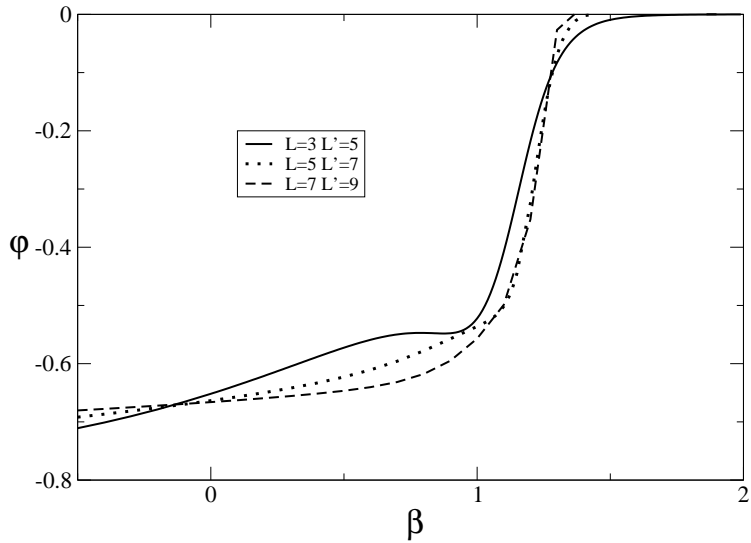}
\end{center}
\caption{$\varphi_{L,L+2}(\beta)$ plotted along the low-$\kappa$-transition line. Again we see crossings corresponding to the self-avoiding walk fixed point, and the collapse transition. 
For the $\Theta$ point we would expect a value of $\varphi=-1/4$. From the above figure the estimate of $\varphi$ is closer to $-0.11$ ($\nu\approx 0.53$), although the value of $\varphi$ does decrease as $L$ is increased. Again with just two solutions, it is not possible to extrapolate.}\label{phiplots}
\end{figure}

 We cross-check these results using additional estimates for the transition points. For the collapse transition we used
   phenomenological renormalisation using
  $\rho$ (see~(\ref{rhoscale}--\ref{phisol})), whilst for the crystallisation transition we looked for solutions to $\lambda_1=\lambda_2=1$.  Plots of $\varphi_{L,L+2}$ are shown in figure~\ref{phiplots} and the estimates for the transition points using these two methods are, again, 
  shown in tables~\ref{values} and~\ref{valuesb}. The estimates for $\nu$  found using the phenomenological renormalisation group are again consistent with a $\Theta$-point transition.

\begin{table}

\caption{Different estimates for the fugacity at the collapse transition, $\kappa_{coll}$, using the different methods described in the text.}\label{values}

\begin{tabular}{@{}lllllll}
\br
L&Nightingale&$\varphi$ 
&$\chi_\rho$&$\chi_{\delta\rho}$& $C_V$ & $\lambda_1=\lambda_2=1$\\
\mr
3& 0.268701&0.258766&0.37705&0.30406&0.31415& --- \\
4&0.280168 &0.270168&0.34815&0.35443& 0.35781 &0.303687 \\
5&0.277469& 0.267450&0.31116&0.26744& 0.28073& --- \\
6& --- &---&0.27017&0.30231& 0.31440 &0.277016\\
7&---& --- &0.28871&0.25980& 0.27225& ---\\
8&---&---&0.28516&0.27886& 0.29573 &0.269245\\
9&---&---&0.28002 &0.25872&0.26949&---\\
\mr
$\infty$&&&$0.271\pm0.003$& $0.258\pm0.001$ & $0.267\pm0.001$&\\
\br
\end{tabular}
\end{table}

\begin{table}
\caption{Different estimates for the value of $\beta$ at the collapse transition, $\beta_{coll}$, using the different methods discussed in the text.}\label{valuesb}
\begin{tabular}{@{}lllllll}
\br
L&Nightingale&$\varphi$&$\chi_\rho$&$\chi_{\delta\rho}$ & $C_V$&$\lambda_1=\lambda_2=1$\\
\mr
3& 1.283504&1.283169& 0.67946&1.08935 & 1.04256&---\\
4&1.205824&1.205822& 0.77646&0.94286& 0.93014 &1.133416\\
5&1.240191&1.240300&  0.98783&1.24091& 1.17257&---\\
6&---&---& 1.03850& 1.13837& 1.05522 &1.220231\\
7&---&---&  1.11407&1.27827& 1.21309&---\\
8&---&---&1.13936& 1.20315& 1.12213 &1.247185\\
9&---&---&1.16533&1.28481 &1.22818&---\\
\mr
$\infty$&&&$1.23\pm0.02$&$1.29\pm0.01$&$1.24\pm0.01$&\\
\br
\end{tabular}

\end{table}

\section{Discussion}

There are two basic phase diagrams proposed in the literature for this model, both based on extended mean-field type calculations. By the mean-field nature of the calculations, they can only propose
general features, and not details of critical behaviour. This is amplified in walk models, since the connectivities, and thus the single walk nature of the model, cannot be taken into account in the local description inherent in the methods used. However, our transfer matrix calculations tend to
confirm the qualitative predictions made by Buzano and Pretti\cite{Buzano:2002hc} rather than those made by \cite{Stilck:1996rq,Serra:2004qu}. In summary, we propose the phase diagram shown qualitatively in figure~\ref{pdfinal}. 
Moving along the self-avoiding walk line, the walk first collapses by the standard $\Theta$-point type 
transition, followed by a further transition where the corners are ejected from the walk, and the walk lines up with one of the two lattice directions. This latter transition is similar to the transition seen in the so called F-model of a gas of semi-rigid self-avoiding loops filling the lattice with a density equal to one ($\kappa\to\infty$).  
In the F-model the transition is in the BKT universality class. We have found evidence that the bond-interacting self-avoiding walk model also displays a BKT transition, but unusually this transition is extended into a line of transitions in the high-$\kappa$ phase.

Whilst the $\Theta$-point model does not show a dense walk phase transition, other models do,
notably the Hydrogen-bonding self-avoiding walk model\cite{Foster:2001xd}, the Bl\"ote-Nienhuis $O(n)$ walk model\cite{Blote:1989rr}
and the $\Theta$-point model with the inclusion of a bending interaction\cite{Buzano:2002hc,Lise:1998fj,Bastolla:qr,Doniach:1996cg,Doye:1998gg}. These models all have one feature in common: they include interactions which are geometrically frustrated. It is expected that this frustration is seen when the density of the walk becomes finite, where the fractal (or Hausdorff) dimension is equal to the lattice dimension\cite{Foster:2003sh}. In the bond-interacting model the exponent found for the collapse transition is clearly larger than $1/2$, hence the Hausdorff dimension is lower than the lattice dimension, and we could therefore expect the critical behaviour not to be modified by any lattice effects. In this case it is reasonable to expect the collapse transition to be in the $\Theta$ universality class.

It is interesting to note that in the three models in which the dense-phase transition line has been studied 
beyond mean-field, i.e. the Bl\"ote-Nienhuis $O(n)$ model, the Hydrogen-bonding model, and the bond-interacting model, all seem to have different classes of transition: 
for the Bl\"ote-Nienhuis $O(n)$
model the
dense-phase transition line was in the Ising universality class\cite{Guo:1999rq}, in the Hydrogen model the transition was critical with a divergent specific heat with a value of $\nu<1$\cite{Foster:2003sh} and now the bond-interacting model with
a BKT transition. 
At first sight it seems as if there are as many different thermodynamic behaviours as 
models one might define, but
preliminary results for a Hydrogen-bonding self-avoiding walk, extended to include solvent quality effects (in the language of a polymer model) indicate that this model displays at least the latter
two behaviours. 

Whilst transfer matrix calculations enable an exploration of the whole phase space, not accessible to Monte-Carlo simulation, the method is hampered by the small number of lattice widths available. Recently we introduced an extension to the CTMRG method for walk models\cite{Foster:2003sh,Foster:2003mb}, but unfortunately this method is not easy to implement for bond-interactions.
 Monte-Carlo simulation may be useful to study the limit $N\to \infty$, enabling an independent study of the different transitions proposed.

\section*{References}
\providecommand{\newblock}{}

\end{document}